\def\rfr#1{eq. (\ref{#1})}

\def\derp#1#2{\rp{\partial{#1}}{\partial{#2}}}

\def\wol{Wolszczan}
\def\kon{Konacki}
\def\mac{Maciejewski}

\def\PSR{PSR B1257+12}

\def\bar{\begin{eqnarray}}
\def\ear{\end{eqnarray}}
\def\bb{\bibitem}
\def\eqi{\begin{equation}}
\def\eqf{\end{equation}}
\def\eqia{\begin{eqnarray}}
\def\eqfa{\end{eqnarray}}
\def\rp#1#2{{#1\over#2}}
\def\ct#1{\cite{#1}}
\def\lb#1{\label{#1}}

\def\oc2{$\mathcal{O}(c^{-2})$}

\documentclass[11pt]{article}
\usepackage{amsmath,amsthm,amscd,amssymb}
\usepackage{latexsym}
\usepackage{graphicx,epsfig}
\usepackage{lscape,rotating}
\begin{document}

\noindent{\bf \LARGE{Constraining the relative inclinations of the planets B and C of the millisecond pulsar \PSR}}
\\
\\
\\
{Lorenzo Iorio}\\
{\it Ministero dell'Istruzione, dell'Universit\`{a} e della Ricerca (M.I.U.R.), Fellow of the Royal Astronomical Society (F.R.A.S.)\\
Address for correspondence: Viale Unit$\grave{a}$ di Italia 68, 70125\\Bari (BA), Italy
\\tel. 0039 328 6128815
\\e-mail: lorenzo.iorio@libero.it}

\begin{abstract}
We investigate on the relative
inclination of the planets B and C orbiting the pulsar PSR B1257+12.  First, we show that the third Kepler law does represent an adequate model for the orbital periods $P$ of the planets, because other Newtonian and Einsteinian corrections  are orders of magnitude smaller than the accuracy in measuring $P_{\rm B/C}$. Then,   on the
basis of available timing data,
we determine the ratio $\sin i_{\rm C}/\sin i_{\rm B}=0.92\pm 0.05$ of the orbital inclinations $i_{\rm B}$ and $i_{\rm C}$ independently of the pulsar's mass $M$.   It turns out that coplanarity of the orbits of B and C would imply a violation of the equivalence principle.
Adopting a pulsar mass range  $1\lesssim M\lesssim 3$, in solar masses (supported by present-day theoretical and observational bounds for pulsar's masses), both face-on
and edge-on
orbital configurations for the orbits of the two planets are ruled out; the
acceptable inclinations for B span the range  36 deg $\lesssim i_{\rm B} \lesssim 66$ deg, with a corresponding relative inclination range
6 deg $\lesssim (i_{\rm C}-i_{\rm B}) \lesssim 13$ deg.

 \end{abstract}


The  6.2-ms \PSR\ pulsar was discovered in 1990 during a high Galactic latitude
search for millisecond pulsars with the Arecibo radiotelescope at
430 Hz \ct{Wol90}. Two years later, \PSR\ turned out to be
orbited by at least two Earth-sized planets--B and C--along almost circular paths \ct{Wol92}. In 1994 it was announced the
discovery of a third, Moon-sized planet--A--in an inner, circular
orbit \ct{Wol94}. Its presence, questioned by Scherer et al. \ct{Sch97}, was
subsequently confirmed in \ct{Kon99, Wol00}.
The relevant orbital parameters of the \PSR\ system are
listed in Table \ref{tavola}.
\begin{table}
\caption{ Relevant parameters \ct{Kon03} of the three planets \ct{Wol92, Wol94}
A, B and C, hosted in the \PSR\ system \ct{Wol90}, derived from analysis of timing data ranging 12 years (1990-2003) collected at the 305-m Arecibo telescope. $P$ is the orbital period, $x$ is the projected barycentric semimajor axis of pulsar's motion and $\gamma\doteq m/M$ \ct{Kon00} is the ratio of the planet's mass to the pulsar's mass. Figures in parentheses are the formal $1\sigma$ uncertainties in the last digits quoted.} \label{tavola}

\begin{tabular}{llll} \noalign{\hrule height 1.5pt}

Planet & $P$ (d) & $x$ (ms) & $\gamma$ $(10^{-6})$\\
\hline
A  & $25.262(3)$ & $0.0030(1)$ & $-$\\
B  & $66.5419(1)$ & $1.3106(1)$ & $9.2(4)$\\
C  & $98.2114(2)$ & $1.4134(2)$ & $8.3(4)$\\
\hline

\noalign{\hrule height 1.5pt}
\end{tabular}
\end{table}
Note that the ratios $\gamma_{\rm B,C}$  of the masses of B and C to $M$ were measured from timing data exploiting their mutual gravitational perturbations \ct{Kon00}, without using the standard reference value $M=\overline{M}\doteq 1.4$M$_{\odot}$ for the pulsar's mass.

Given the peculiarity of the \PSR\ system, it is certainly important to deepen the knowledge of the orbital configuration of its planets in order to gain insights about the evolutionary dynamics of such a rare system.
Here, without making any a priori assumptions about
the  inclinations of B and C,
we wish to constrain them
from the available timing data by assuming a reasonable interval of masses for the hosting pulsar.

To this aim, we will exploit the third Kepler law whose use is justified in detail below.
 First of all, let us note that, by defining $s\doteq \sin i$, it is possible to write the planetary relative (i.e. pulsar-to-planet) semimajor axis $a$ in terms of the measured quantities $\gamma$ and $x$ as
\eqi a =\left(1+\rp{M}{m}\right)\rp{xc}{s}=\left(\rp{1+\gamma}{\gamma}\right)\rp{xc}{s},\lb{semia}\eqf where $c$ is the speed of light in vacuum. Note that both $x$  and $\gamma$  were phenomenologically determined
in \ct{Kon03} independently of the third Kepler law itself and of the pulsar's mass. The third Kepler's law is \eqi\left(\rp{P}{2\pi}\right)^2=\rp{a^3}{GM(1+\gamma)},\lb{3kpa}\eqf from which it is possible to express the pulsar's mass in terms of the phenomenologically determined quantities $x,\gamma, P$, apart from $s$ which will be considered as unknown.

 Note that a purely Keplerian model for the orbital period is quite adequate because non-Keplerian corrections like those due to the oblateness of the pulsar (if any), to the planet-planet interaction and to the 1PN $\mathcal{O}(c^{-2})$ terms are negligible given the present-day accuracy in determining $P_{\rm B/C}$.
 Indeed, concerning the oblateness of the central mass, its contribution $\Delta P^{(\rm obl)}$ to the orbital period of an orbiting test particle can be written as   \cite{Ior05}
 \eqi \Delta P^{(\rm obl)}=-\rp{6\pi R^2 J_2}{\sqrt{GMa}},\eqf where $|J_2|<1$ represents the first even zonal harmonic coefficient of the multipolar expansion of the gravitational potential of the pulsar\footnote{In this rough order-of-magnitude estimate there is no need to take into account contributions due to the peculiarity of the matter state in the neutron star.} and $R$ is the pulsar's radius; by assuming typical values
 \footnote{As a consequence, we also used $i_{\rm B}=53$ deg and $i_{\rm C}=47$ deg obtained in \cite{Kon03} with $\overline{M}$.}
 $R=\overline{R}\doteq 10$ km and $M=\overline{M}=1.4$M$_{\odot}$ one has
 \eqi \Delta P^{(\rm obl)}_{\rm C/B}\approx -J_2\times 10^{-12}\ {\rm d}.\eqf
 The corrections $\Delta P^{(\rm 3rd\ body)}$ to the orbital period of a planet of mass $m$ induced by another planet of mass $m^{'}$ can be written as \cite{Ior05}
 \eqi \Delta P^{(\rm 3rd\ body)} = -\rp{4\pi Gm^{'}}{n {a^{'}}^3},\eqf where $n=\sqrt{Gm/a^3}$ is the Keplerian mean motion of the perturbed planet and $a^{'}$ is the perturber's semimajor axis. In the case of the planets B and C it turns out that
 \eqi \Delta P^{(\rm 3rd\ body)}\approx 10^{-15}\ {\rm d}.\eqf
 The fact that the planets B and C are in a 3:2 resonance  does not affect their orbital periods. Indeed, according to the Lagrange's perturbation equation for the variation of $a$ \cite{Clem}, \eqi \dot a \propto\derp{H_1}{\sigma},\eqf where $\sigma=-nT_p$ is related to the time of pericentre's passage $T_p$ and $H_1$ is the interacting Hamiltonian  of eq.(23) in ref. \cite{Kon00}. Since $H_1$ does not explicitly contain  $\sigma$ there is no secular change in the semimajor axis of B and C. The mutual perturbing effects employed in ref. \cite{Kon00} to estimate $\gamma_{\rm B}$ and $\gamma_{\rm C}$  are not the corrections to the Keplerian orbital periods.

 The 1PN Post-Newtonian correction $\Delta P^{(\rm 1PN)}$ to the orbital period of order $\mathcal{O}(c^{-2})$ can be written as  \cite{Dam86}
 \eqi \Delta P^{(\rm 1PN)}=\rp{3\pi}{c^2}\sqrt{GM a};\eqf  it turns out that \eqi \Delta P^{(\rm 1PN)}_{\rm C/B}\approx 10^{-6}\ {\rm d}.\eqf The latter contribution is the most important post-Keplerian correction to $P$, but it is two orders of magnitude smaller than the $1\sigma$ formal errors in the phenomenologically determined orbital periods quoted in Table \ref{tavola}.

Thus, from \rfr{semia}-\rfr{3kpa} we have \eqi GM =\left[\rp{2\pi(1+\gamma)}{P}\right]^2\left(\rp{xc}{\gamma s}\right)^3,\lb{GM}\eqf
Taking the ratio of \rfr{GM} for both B and C allows us to get information about the relative orbit inclination independently of $M$ itself: indeed, we have, from Table \ref{tavola}
\eqi S\doteq \rp{s_{\rm C}}{s_{\rm B}}=\left(\rp{P_{\rm B}}{P_{\rm C}}\right)^{2/3}\left(\rp{x_{\rm C}\gamma_{\rm B}}{x_{\rm B}\gamma_{\rm C}}\right)\left(\rp{1+\gamma_{\rm C}}{1+\gamma_{\rm B}}\right)^{2/3}=0.92\pm 0.05.\lb{seni}\eqf
The $1\sigma$ error was conservatively  assessed by propagating through \rfr{seni} the uncertainties in $P_{\rm B},P_{\rm C},x_{\rm B},x_{\rm C},\gamma_{\rm B},\gamma_{\rm C}$ quoted in Table \ref{tavola}  and linearly adding the resulting biased terms.
It turns out that the most important sources of errors are $\gamma_{\rm B}$ and $\gamma_{\rm C}$ yielding $\delta S_{{\gamma_{\rm B}}}=0.02$ and   $\delta S_{{\gamma_{\rm C}}}=0.03$.

It maybe interesting to note that, by assuming $\sin i_{\rm B}=\sin i_{\rm C}$, the quantity $S$ in \rfr{seni} may be interpreted  as a measure of a violation of the equivalence principle. Indeed, according to \cite{Nob08}, by putting $m_{\rm g}=m_{\rm i}(1+\eta)$ for the gravitational and inertial masses of a test particle orbiting a central body of mass $M$, the $2-$body problem encompassing a violation of the equivalence principle can be
described by the same formulas derived in the classical $2-$body problem, provided that whenever they contain the
product $GM$, we substitute it with $GM(1 + \eta)$. Applying it to \rfr{GM} written for the planets B and C, it turns out that $S$ can be interpreted as \eqi S \doteq\left(\rp{1+\eta_{\rm C}}{1+\eta_{\rm B}}\right)^{1/3}.\eqf Thus, coplanarity and the result of \rfr{seni} would yield a violation of the equivalence principle in the \PSR\ system at $1.6\sigma$ level. However, it must be noted that such a test would be
stronger if one had evidence that the composition of the two planets was very different.

\begin{figure}
\begin{center}
\includegraphics[width=\columnwidth]{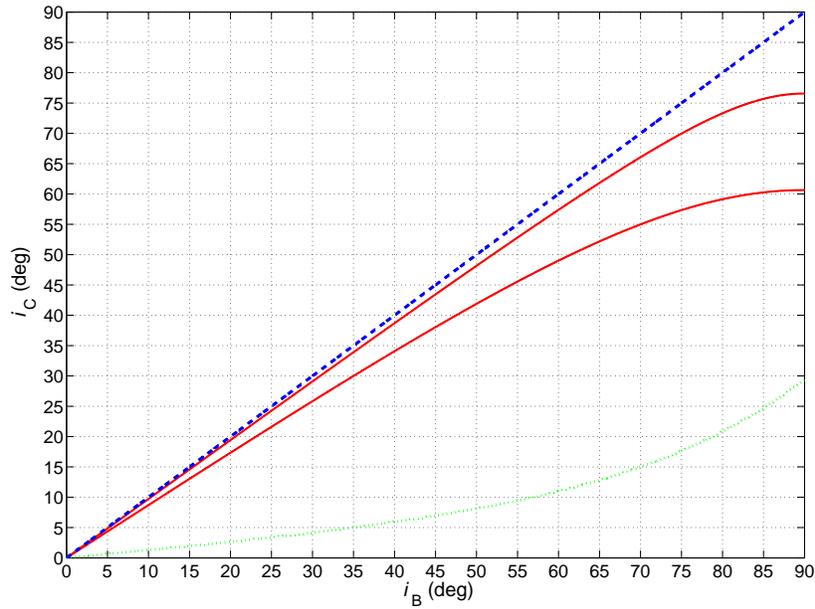}
\end{center}
\caption{\label{inclinazione}  Plot of $i_{\rm C}=\arcsin(S\sin i_{\rm B})$ for the minimum (lower red curve) and maximum (upper red curve) values of $S$ according to \rfr{seni}. Such constraints are independent of the pulsar's mass. The blue dashed line represents the coplanarity case. The distance between the blue coplanarity line and the lower red curve yields the maximum value of the relative inclination $I$ allowed by the system's parameters along with their uncertainties; it is depicted in green and shows that $I\gtrsim 10$ deg for $i_{\rm B}\gtrsim 56$ deg.}
\end{figure}

The constrain of \rfr{seni} is a consistent, genuine dynamical one which does not make use of any assumption about $M$; the authors of \cite{Kon03} did not obtain it.
However, we note that the values of $\gamma_{\rm B/C},$ entering \rfr{seni}, were measured in \ct{Kon03} by using the model of \ct{Kon00} which neglects terms in $\sin^2(I/2)$, where $I=|i_{\rm B}-i_{\rm C}|$ is the relative inclination of the orbital planes of B and C assumed to be $I\lesssim 10$ deg. In Figure \ref{inclinazione} it is plotted the allowed region in the plane $\{i_{\rm B},i_{\rm C}\}$, according to \rfr{seni}, delimited by the minimum and maximum values of the ratio $S$.  The blue dashed line represents the coplanarity condition, while the green line is the maximum  value of $I$ allowed by the system's parameters.
It can be noted that for face-on ($i\rightarrow 0$ deg) geometries the orbital planes tend to be coplanar. Most remarkable deviations from coplanarity (more than 10 deg with a maximum of about 32 deg for $i_{\rm B}=90$ deg) occur for edge-on ($i\rightarrow 90$ deg) geometries, but in such cases caution is required since we would fall outside the $I\lesssim 10$ deg condition on which our analysis relies upon; Figure \ref{inclinazione} shows that this occurs for $i_{\rm B}\gtrsim 56$ deg.
In the following we will explore the  viability of   various possible orbital inclinations in
terms of physically plausible values of  $M$.

%
%

We will study the behavior of \rfr{GM} for B and C as a function of $i_{\rm B}$ (because of \rfr{seni}) in order to constrain the inclinations. Indeed, since we have no independent information at all on $i_{\rm B/C}$, we will use a reasonable interval of masses for the pulsar to constrain them (and their relative inclination $I$ through \rfr{seni} which is independent of $M$).  We wish to preliminarily notice that, in principle, PSR B1257+12, as a member of the rare class of the planetary pulsars, may have had a different formation and evolution with respect to the other neutron stars. However, in absence of any other indication on the details for the evolutionary history of PSR B1257+12, we will  rely the following analysis upon standard mass intervals, commonly adopted for other kinds of neutron stars.  This is a notable difference with respect to \cite{Kon03} in which the pulsar's mass was kept fixed to $1.4$M$_{\odot}$.
Theoretically speaking, different Equations-Of-State  for nuclear matter inside  neutron star yield different pulsar's mass ranges; we will adopt $1-3$M$_{\odot}$ \cite{Lat04}. Let us note that
present-day
observations are all compatible with that range.
As to the lower bound, all the best determinations of the
mass of a neutron star fall well above 1 M$_{\odot}$, approaching that value only for the most uncertain
cases (see ref. \cite{Sta06} for an overview of measured pulsar masses). As to the upper bound, the highest
securely measured value of the mass of a pulsar is that recently obtained for PSR J04374715
(about $1.76\pm 0.20$ M$_{\odot}$ \cite{Verb08}). Anyway, our adopted mass range also includes the
cases of more massive neutron stars (e.g. ref. \cite{Fr08}), whose claimed high
mass must still be confirmed by additional investigations.

\begin{figure}
\begin{center}
\includegraphics[width=\columnwidth]{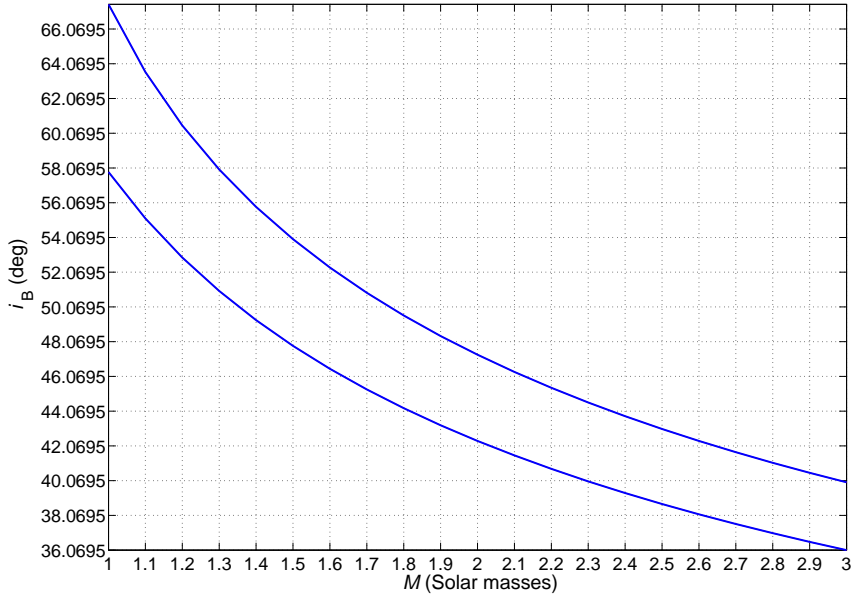}
\end{center}
\caption{\label{figupul}  Allowed region for $i_{\rm B}$ as a function of $M$. The $1-3$M$_{\odot}$ interval of \cite{Lat04} has been chosen.}
\end{figure}

The $1\sigma$ error in the value of $M$ calculated using \rfr{GM} can be conservatively evaluated by propagating the uncertainties  in $P,x,\gamma$
of Table \ref{tavola}, and linearly summing the resulting individual biased terms. It is of the order of $13\%$  and the major contribution to it turns out to be due to $\gamma$.
By considering the allowed regions for $M$ determined by the curves $M\pm\delta M$ by means of \rfr{GM} applied to both B and C, it turns out that
the tightest constraints come from B whose allowed region is entirely enclosed in that due to C.
In Figure \ref{figupul} we depict the constraints on $i_{\rm B}$ for $1 \lesssim M \lesssim 3$ in solar masses.
As can be noted, it turns out that
$36\lesssim i_{\rm B}\lesssim 66$ deg.   From an inspection of Figure \ref{inclinazione} it turns out that the relative inclination $I$ is different from zero being $6 \lesssim I\lesssim 13$ deg. It is interesting to note that larger values of $I$, which, at least to a certain extent, may  still be compatible with the analysis presented here\footnote{After all, $\sin^2 I/2=0.007$ for $I=10$ deg; a relative inclination of, say, $I = 15$ deg would yield $\sin^2 I/2=0.02$.}, are ruled out by the lower bound on the pulsar's mass.

\begin{figure}
\begin{center}
\includegraphics[width=\columnwidth]{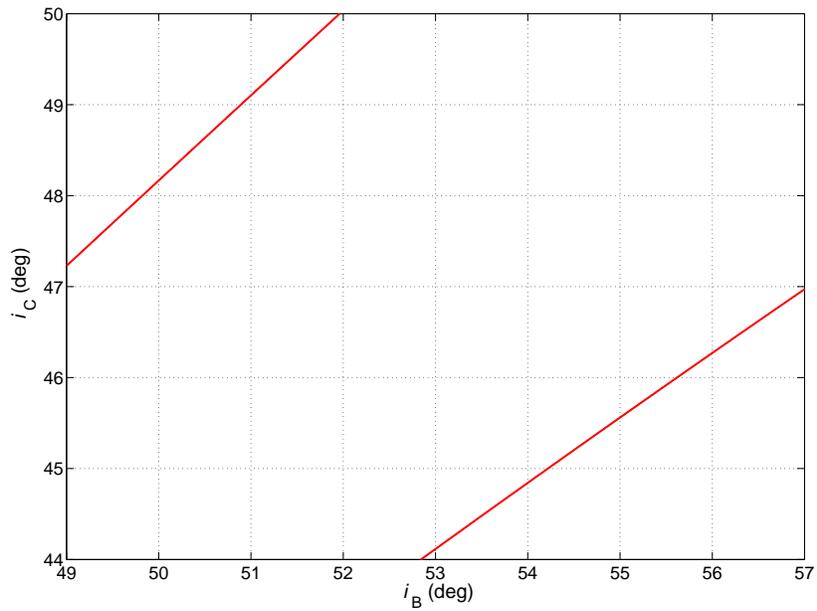}
\end{center}
\caption{\label{dettaglio}  Detail of Figure \ref{inclinazione}. Zoomed section of Figure \ref{inclinazione}, focusing on the most likely region for $i_{\rm B}$ and $i_{\rm C}$ ($49$ deg $\leq i_{\rm B}\leq 57$ deg)  $\times$ ($44$ deg $\leq i_{\rm C}\leq 50$ deg) obtained in \cite{Kon03} under the assumption $M=1.4$M$_{\odot}$. It is possible to see that only a portion of the region, delimited by the red lines, is compatible with the mass-independent constrain of  \rfr{seni}.}
\end{figure}

The authors of \cite{Kon03}, by using $M=\overline{M}$, obtain $49$ deg $\leq i_{\rm B}\leq 57$ deg and $44$ deg $\leq i_{\rm C}\leq 50$ deg.
On one hand, our analysis confirms-as expected-that a larger range of values for $i_{\rm B}$ and $i_{\rm C}$ is allowed when a suitable interval of values for the mass of the pulsar is taken into account. On another hand, thanks to the new constrain implied by \rfr{GM},   it is now possible to show that not all the combinations of orbital inclinations $i_{\rm B}$ and $i_{\rm C}$ indicated in ref. \cite{Kon03} are acceptable (even when adopting the same uncertainty level of ref. \cite{Kon03}). In fact, as depicted in Figure \ref{dettaglio},  the rectangle of acceptable values  ($49$ deg $\leq i_{\rm B}\leq 57$ deg)  $\times$ ($44$ deg $\leq i_{\rm C}\leq 50$ deg) indicated in ref. \cite{Kon03} is not entirely included in the region allowed by the mass-independent constrain of \rfr{seni}.
%
%
%
%
%


\section*{Acknowledgments}
I thank A. Possenti for his useful criticisms, suggestions and contributions which greatly contributed to improve the manuscript.


\end{document}